\documentclass[proof]{WileyASNA-v1}

\articletype{Article Type}%

\received{26 April 2016}
\revised{6 June 2016}
\accepted{6 June 2016}

\usepackage{makecell}

\begin{document}

\title{An X-ray View of the Hot Circum-Galactic Medium\protect\thanks{An X-ray View of the Hot Circum-Galactic Medium}}

\author[1]{Jiang-Tao Li$^{1,*}$}

\authormark{Li \textsc{et al}}

\address[1]{\orgdiv{Department of Astronomy}, \orgname{University of Michigan}, \orgaddress{\state{MI}, \country{U.S.A.}}}

\corres{*Jiang-Tao Li, 311 West Hall, 1085 S. University Ave, Ann Arbor, MI, 48109-1107. \email{pandataotao@gmail.com}}

\presentaddress{311 West Hall, 1085 S. University Ave, Ann Arbor, MI, 48109-1107}

\abstract{The hot circum-galactic medium (CGM) represents the hot gas distributed beyond the stellar content of the galaxies while typically within their dark matter halos. It serves as a depository of energy and metal-enriched materials from galactic feedback and a reservoir from which the galaxy acquires fuels to form stars. It thus plays a critical role in the coevolution of galaxies and their environments. X-rays are one of the best ways to trace the hot CGM. I will briefly review what we have learned about the hot CGM based on X-ray observations over the past two decades, and what we still do not know. I will also briefly prospect what may be the foreseeable breakthrough in the next one or two decades with future X-ray missions.}


\maketitle

\footnotetext{\textbf{Abbreviations:} X-ray View of the hot CGM}


\section{What we already know about the hot CGM?}\label{sec:background}

\begin{figure*}[t]
\centerline{\includegraphics[width=0.98\textwidth]{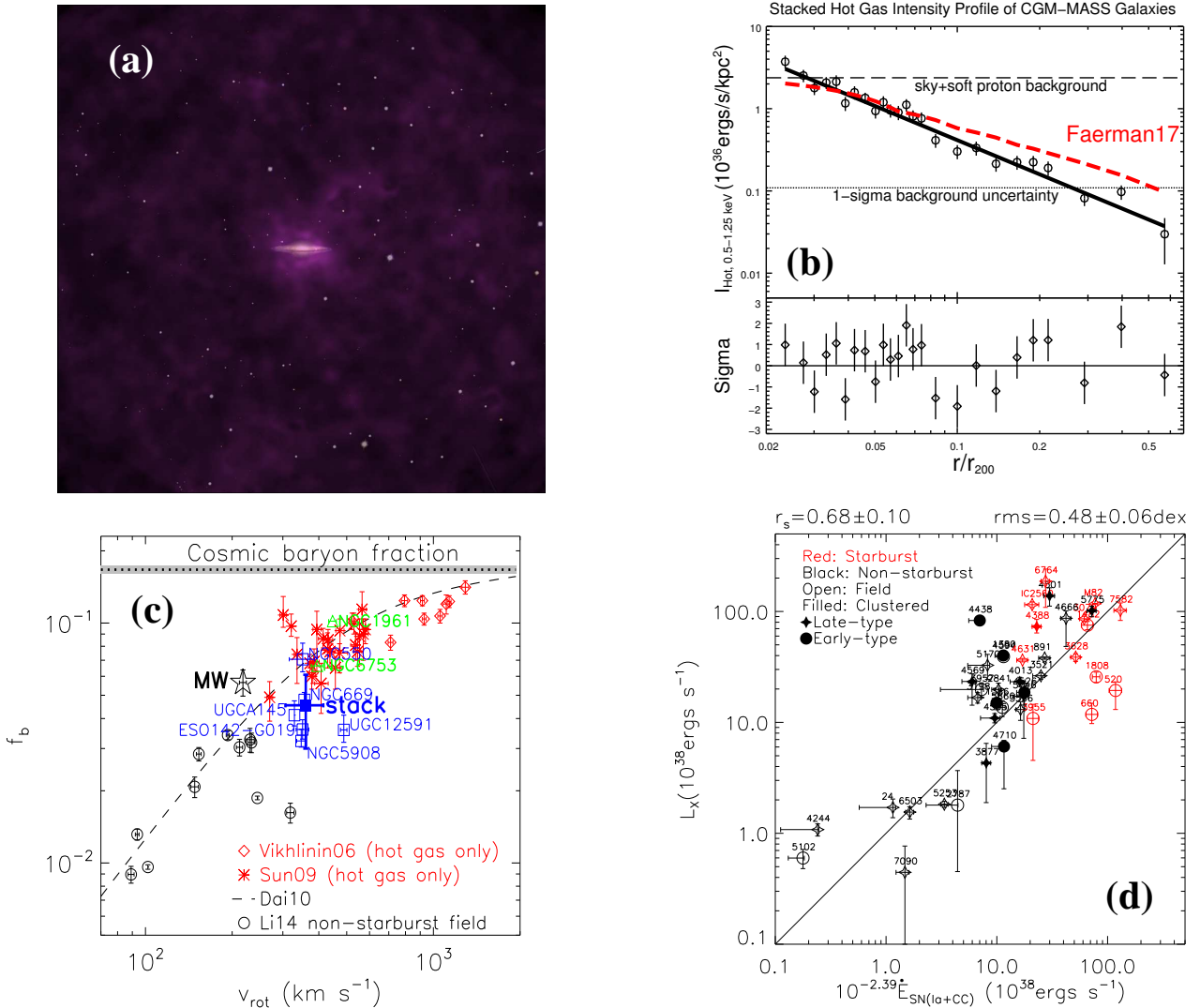}}
\vspace{-0.3in}
\caption{\textbf{(a)} Stacked diffuse soft X-ray images of the CGM-MASS galaxies (purple; from ESA press release, with original data from \citealt{Li18}) overlaid on the SDSS image of NGC~5908 (white). \textbf{(b)} Stacked radial intensity profiles of the hot gas component of the CGM-MASS galaxies \citep{Li18}. The radial distance of different galaxies has been rescaled to $r_{\rm 200}$. The solid line is the best-fit $\beta$-function. The dashed and dotted lines show the sky+soft proton background and the 1$\sigma$ uncertainty. The red dashed curve is the \citet{Faerman17} model scaled with $r_{\rm 200}$ of a MW-sized halo. \textbf{(c)} Baryon fraction ($f_{\rm b}$) v.s. rotation velocity ($v_{\rm rot}$) \citep{Li18}. The cosmic baryon fraction (dotted line) with error (shaded area), a fitted relation from \citet{Dai10} (dashed), the MW (\citealt{Miller15}; black star), samples of non-starburst field spirals (\citealt{Li14}; black circles), galaxy groups (\citealt{Sun09}; red stars), and galaxy clusters (\citealt{Vikhlinin06}; red diamonds) are plotted for comparison. \textbf{(d)} $L_{\rm X}$ v.s. the total SN energy injection rate $\dot{E}_{\rm SN(CC+Ia)}$ fitted with a linear relation \citep{Li13b}.}\label{fig01}
\end{figure*}

The hot circum-galactic medium (CGM), or sometimes named as the galactic corona, represents X-ray emitting hot gas typically at a temperature of $kT\sim10^{6-8}\rm~K$ around galaxies. The hot CGM could extend beyond the galaxy stellar disk and bulge and typically out to the virial radius of the dark matter halo or even beyond (Fig.~\ref{fig01}a\textbf{,b}). In principle, the hot CGM could cool and fall back onto the galaxy, providing fresh fuel to continue star formation. It can also be ejected out of the galaxy, entraining a lot of energy and metal-enriched material from galactic stellar (stellar wind, supernovae: SNe, etc.) or AGN feedback.

\subsection{Key science related to the hot CGM}\label{subsec:KeyScience}

As probably the most spatially extended and volume filling phase of the CGM, the hot gas could be important in the co-evolution of galaxies and their environments, in particular, in the mass and energy budgets of the galaxies.

The hot CGM could contain a significant fraction of the ``missing baryons'' (e.g., \citealt{Anderson10,Bregman07,Bregman18}). As the soft X-ray emission from hot gas is dominated by metal lines whose strength is proportional to both the gas density and metallicity, the detected soft X-ray emission around nearby galaxies is largely biased to the highly metal enriched CGM around actively star forming (SF) galaxies. This part dominates the X-ray emission but may contain only a small fraction of the baryon mass (e.g., \citealt{Crain13}). In order to measure the total mass contained in the extended hot CGM, we need to detect the hot gas at large galacto-centric radii where the feedback material encounters the metal-poor gas accreted from outside the galaxies (e.g., \citealt{Benson00,Toft02}). Since $L^\star$ [an $L^\star$ galaxy has a roughly comparable optical luminosity as the Milky Way (MW)] or sub-$L^\star$ galaxies are not massive enough to gravitationally heat the gas to X-ray emitting temperature, while massive elliptical galaxies are often located in clusters, the best candidates to search for the accreted and gravitationally heated hot CGM would be massive disk galaxies. There are many efforts searching for the extended hot CGM around such galaxies (e.g., \citealt{Rasmussen09,Anderson11,Anderson13,Dai12,Bogdan13,Bogdan15,Bogdan17,Li16b,Li17}), while the latest efforts indicates that even in these most massive and isolated systems, the hot CGM may not be sufficient to account for the ``missing baryons'' (Fig.~\ref{fig01}c; \citealt{Li18}). 

The hot CGM could also contain a significant fraction of the energy injected by various types of galactic feedback (e.g., \citealt{Li13b}). This gas phase could contain energy at least in a few forms such as radiative loss, kinetic energy of global motion, and turbulent energy. We herein consider the global energy budget of galaxies with a few simplified assumptions. We do not consider {\sl AGN feedback} as it is the most important in very massive systems and on large scales (e.g., for the intracluster medium ICM), and may not have continuous energy injection over the lifetime of galaxies. Most of the CGM phases are detected typically at $r<20\rm~kpc$ (e.g., \citealt{Li13a}), where {\sl gravitational heating} is typically not as important as feedback. As most of the energy input from {\sl young stars} (radiation, stellar wind, etc.) is reprocessed by the ISM locally near SF regions and the CGM is optically thin to the reprocessed emission (peaked in infrared), stars do not contribute much to the global energy balance of the CGM. Therefore, in the simplest case, we only consider {\sl energy injection from SNe} and compare it to the {\sl energy detected in various CGM phases}. Table~\ref{tab:EnergyBudget} summarizes a very rough estimate of the SNe energy budget in the CGM. We emphasize that reliable observations of many CGM phases are still very limited, so the numbers in Table~\ref{tab:EnergyBudget} are just order of magnitude estimate with very different samples (many are just case studies) and often highly uncertain assumptions. The basic conclusion is that a significant fraction of SNe energy is not detected around galaxies, which is often regarded as the ``missing galactic feedback'' problem (e.g., \citealt{Wang10}). Such a problem is also revealed by the lack of heavy elements from feedback in the CGM (e.g., \citealt{Wang10}). 

\begin{center}
\begin{table}[t]%
\centering
\caption{Measured SNe Energy Budget in the CGM.\label{tab:EnergyBudget}}%
\tabcolsep=0.pt%
\begin{tabular*}{20pc}{@{\extracolsep\fill}ccc@{\extracolsep\fill}}
\toprule
\textbf{Phase} & \textbf{Fraction}  & \textbf{Reference} \\
\midrule
\makecell{Thermal energy of hot\\gas radiated in X-ray} & $\sim1\%\dot{E}_{\rm SN}$ & \citet{Li13b} \\
\makecell{Kinetic energy of the\\global motion of hot gas} & $\lesssim1\%\dot{E}_{\rm SN}$ & \citet{Li13b}\tnote{$^1$} \\
\makecell{Turbulent energy of small\\-scale motion of hot gas} & $\lesssim0.1\%\dot{E}_{\rm SN}$ & \citet{Hitomi16}\tnote{$^2$} \\
\makecell{Kinetic energy of cold\\ gas outflow} & $<1\%\dot{E}_{\rm SN}$ & \citet{Contursi13} \\
\makecell{Radiative cooling\\of cool gas} & $\lesssim5\%\dot{E}_{\rm SN}$ & \citet{Otte03} \\
\makecell{Cosmic ray} & $\sim5\%\dot{E}_{\rm SN}$ & \citet{Li16a}\tnote{$^3$} \\
\makecell{Magnetic field} & $<5\%\dot{E}_{\rm SN}$ & \makecell{\citet{MoraPartiarroyo19}} \\
\bottomrule
Total CGM & $<20\%\dot{E}_{\rm SN}$ & - \\
\bottomrule
\end{tabular*}
\begin{tablenotes}
\item A rough estimate of the fraction of SNe energy injection rate ($\dot{E}_{\rm SN}$) distributed into different phases of the CGM. Other energy sources (AGN, gravitational heating, stellar radiation and wind, etc.) and the energy consumed locally in the ISM are not included here. The estimates are made based on different samples (many are just case studies) and different assumptions.
\item[1. ] Estimated assuming a pressure driven adiabatic expansion.
\item[2. ] Estimated assuming the same turbulent to thermal pressure ratio as in the Perseus cluster.
\item[3. ] Estimated assuming a typical CR hadron-to-lepton energy density ratio of 50 \citep{Strong10}.
\end{tablenotes}
\end{table}
\end{center}


\subsection{Detection and statistics of the extended X-ray emission around nearby galaxies}\label{subsec:DetectionXray}

Extended soft X-ray emission, which is often thought to be largely contributed, if not dominated by hot gas, has been commonly detected around various types of nearby galaxies. The detection, however, is often limited to a few tenth of kpc from the galactic center around isolated galaxies (e.g., \citealt{Strickland04a,Tullmann06a,Li13a,Bogdan15,Kim19}). Apparent scaling relations between the X-ray properties of the ``hot CGM'' and various galaxy parameters have been well established in many existing works (e.g., Fig.~\ref{fig01}d; \citealt{OSullivan03,Strickland04b,Tullmann06b,Boroson11,Kim13,Li13b,Mineo14,Wang16}). Most of these scaling relations indicate that the soft X-ray emissions around galaxies are connected to the stellar feedback activities in the galaxies, but different types of galaxies have clearly different X-ray properties and scaling relations. In general, in addition to core collapsed SNe which are related to the young stellar population, Type~Ia SNe related to the relatively old stellar population could also be important, especially in early-type galaxies (e.g., \citealt{Li15}). On the other hand, gravitational heating, which could be important in massive isolated galaxies, has been suggested but not yet firmly evidenced based on existing observations (e.g., \citealt{Li17}).


\section{What we still don't know?}\label{sec:OpenQuestion}


\subsection{What are the physical, chemical, and dynamical properties of the hot CGM?}\label{subsec:Physics}

\begin{figure*}[t]
\centerline{\includegraphics[width=1\textwidth]{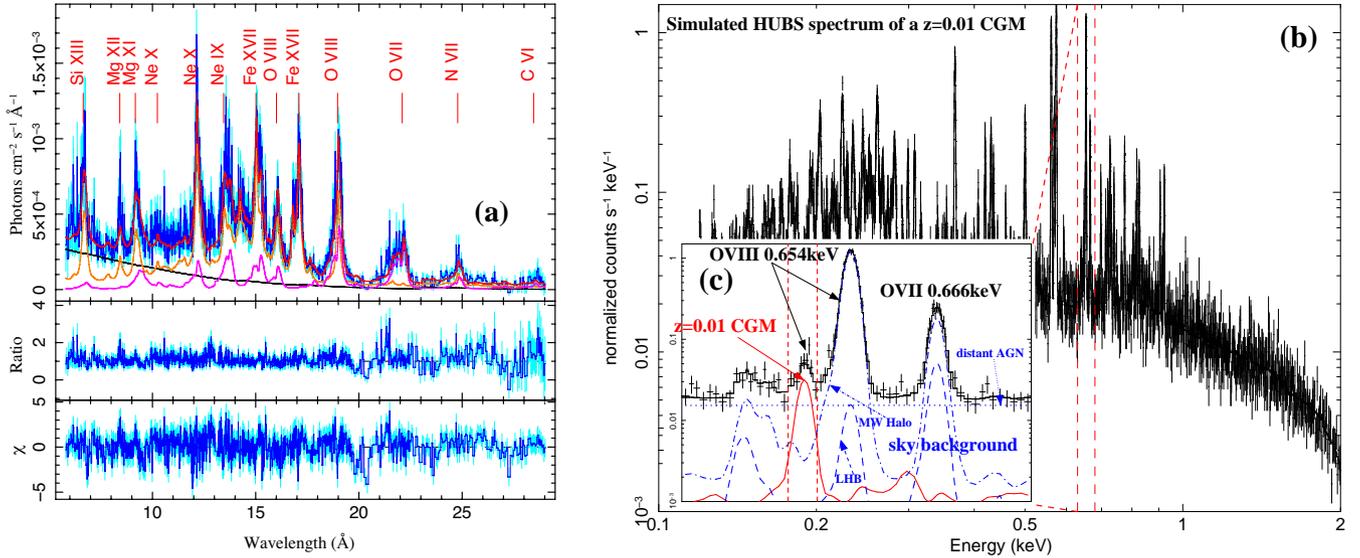}}
\vspace{-0.1in}
\caption{\textbf{(a)} XMM-Newton/RGS spectrum of M82 and the best-fit model [red curve; different components in different colors: power law (black), thermal (orange), CX (purple); \citealt{Zhang14}]. \textbf{(b)} Simulated $1\rm~Ms$ HUBS spectrum of the hot CGM at $r\leq0.2r_{\rm 200}$ around a $z=0.01$ ($d\sim50\rm~Mpc$) galaxy (e.g., NGC~5908, Table~\ref{tab:NarrowBandImaging}; \citealt{Li16b}). \textbf{(c)} Zoom-in of panel~(b) in 0.63-0.68~keV. The red curve is the hot CGM ($0.6\rm~keV$ plasma), while the blue curves are various sky background components (local hot bubble LHB; MW halo; distant AGN). The redshifted OVIII line from the CGM can be separated from the background.}\label{fig02}
\end{figure*}

In addition to thermal plasma, the soft X-ray emission around nearby galaxies could also be contributed by other non-thermal components. One of the most important non-thermal components is charge exchange (CX), which is characterized with a few soft X-ray emission lines (e.g., \citealt{Smith12}) and has been evidenced with high resolution grating spectra of the diffuse soft X-ray emission around a few nearby galaxies (e.g., Fig.~\ref{fig02}a; \citealt{Ranalli08,Liu11,Liu12}). CX could be important or even dominant in some cases. However, in most of the X-ray observations of nearby galaxies, the low resolution CCD spectra prevent us from separating the CX from the thermal plasma component. Even if most of the detected soft X-ray emission is produced by the hot plasma, the thermal structure of gas can be complicated (e.g., \citealt{HodgesKluck18}) and it is often difficult to decompose different gas components (at different temperature and density) with the existing CCD imaging spectroscopy observations (e.g., \citealt{Li13a}).

The importance of CX indicates that there is a strong cool-hot gas interaction in nearby galaxies. A very hot and tenuous gas component therefore highly likely exists and could be produced by the thermalized SNe ejecta (e.g., \citealt{Strickland02,Tang09}). This gas component has too low density and X-ray emissivity, and the best tracer of it is the Fe~K shell emission lines at $6-7\rm~keV$. \citet{Strickland07,Strickland09} claimed the detection of ionized Fe~K line in the nuclear region of M82, which is the most solid such detections so far, but the detection may be highly contaminated by stellar sources (e.g., \citealt{Revnivtsev09}). Therefore, there is still no direct evidence on the existence of such a hot gas component, and its physical and chemical parameters, as well as the related cool-hot gas interactions, are still poorly understood.

Even if we can assume thermal origin of the X-ray emission, the chemical and dynamical properties of the hot CGM are still poorly constrained, based on the low-resolution CCD spectra. It is therefore difficult to measure the amount of heavy elements and kinetic/turbulent energy contained in the hot CGM. Furthermore, because in the modeling of the low-resolution X-ray spectra, the emission measure (or gas density) is degenerated with the metallicity, the measured gas density and mass also have large uncertainties (e.g., \citealt{Bogdan13,Li16b}). These uncertainties strongly affect our understanding of the metal, energy, and baryon budget of the hot CGM.


\subsection{Hot gas distribution at large radii}\label{subsec:RadialDistribution}

Radial distribution of the hot CGM around galaxies is often poorly constrained at large galacto-centric radii because of the low density thus X-ray emissivity of the hot gas. Furthermore, in broad-band imaging observations, the detection limit of faint extended features is limited by the uncertainty of the background, instead of the depth of the observations (Fig.~\ref{fig01}b). Therefore, the hot CGM around individual isolated galaxies, even the most massive ones, is typically detected to $r\lesssim50\rm~kpc$, or $\sim(10-20)\%r_{\rm 200}$ (e.g., \citealt{Dai12,Bogdan13,Li17}). At this radius, the radial distribution of the hot CGM could still be highly affected by galactic feedback. It is therefore difficult to distinguish different models based on the directly measured hot gas radial distribution (Fig.~\ref{fig01}b).

Furthermore, because the origin of gas at larger radii may be quite different (more gas with external instead of internal origin thus lower metallicity), the poor constraint of gas distribution at $r\gtrsim(10-20)\%r_{\rm 200}$ prevents us from both searching for the accreted hot CGM and measuring the total mass and energy contained in the CGM. It is not impossible that the hot CGM extends beyond the virial radius of the dark matter halo thus contains a larger fraction of the missing baryons \citep{Bregman18}, but better observations are needed to constrain the hot gas distribution at large radii.


\begin{center}
\begin{table}[t]%
\centering
\caption{$FoM$ of Some Current and Future X-ray Missions.\label{tab:FoM}}%
\tabcolsep=0pt%
\begin{tabular*}{20pc}{@{\extracolsep\fill}lcccc@{\extracolsep\fill}}
\toprule
\textbf{Mission} & \textbf{$E/\Delta E$}\tnote{$\dagger$}  & \textbf{$A_{\rm eff}/\rm cm^{2}$} & $\Omega_{\rm FOV}/\rm deg^2$ & $FoM$ \\
\midrule
Chandra/ACIS-I & 10 & 110 & 0.079 & 87 \\
XMM-Newton/EPIC\tnote{$\ddagger$} & 15 & 1000 & 0.25 & 3750 \\
Suzaku/XIS & 10 & 1000 & 0.09 & 900 \\
eROSITA & 10 & 1000 & 1.0 & 10000 \\
XRISM & 100 & 70 & 0.0023 & 22 \\
Athena/X-IFU & 240 & 5000 & 0.0069 & 8280 \\
Lynx & 200 & 10000 & 0.0069 & 13800 \\
HUBS & 300 & 500 & 1.0 & 150000 \\
AXIS & 10 & 7000 & 0.05 & 3500 \\
Super-DIOS & 300 & 1000 & 0.25 & 75000 \\
\bottomrule
\end{tabular*}
\begin{tablenotes}
\item $FoM$ of a few current and future focusing X-ray telescopes for emission line detections of extended sources. Parameters are mostly obtained from the website of each mission.
\item[$\dagger$] Energy resolution $E/\Delta E$ and effective area $A_{\rm eff}$ are typical values in soft X-ray ($\sim1\rm~keV$).
\item[$\ddagger$] For a combination of PN + 2 MOS CCDs.
\end{tablenotes}
\end{table}
\end{center}

\section{What can we do in the future?}\label{sec:Future}


\subsection{Future focusing X-ray telescopes}\label{subsec:FutureXMissions}

In the next 10-20 years, there are a few focusing soft X-ray telescopes either already approved or currently in concept study. Most of these telescopes have higher energy resolution and/or better sensitivity than the existing X-ray missions, thus could potentially improve our understanding of the hot CGM. Table~\ref{tab:FoM} summarizes basic parameters of these future X-ray missions in comparison with some existing ones.

For a quantitative comparison among these X-ray missions in the study of the hot CGM, we calculate the Figure-of-Merit ($FoM$) for emission line detections of extended sources for the major instrument of each mission, which is defined as:
\begin{eqnarray}\label{equ:FoM}
FoM=RA_{\rm eff}\Omega_{\rm FOV},
\end{eqnarray}
where $R$ is the energy resolution in $E/\Delta E$, $A_{\rm eff}$ is the effective area in $\rm cm^{2}$, and $\Omega_{\rm FOV}$ is the field of view (FOV) in $\rm deg^2$. The $FoM$ describes the efficiency of an instrument in detecting soft X-ray emission lines from extended sources. 

Many future X-ray missions have significantly improved energy resolution in imaging spectroscopy mode with the micro-calorimeter detector in replacement of CCD (\citealt{McCammon05}; e.g., XRISM, Athena, Lynx, HUBS, and Super-DIOS). The most direct effect is the significantly higher $FoM$ than CCD X-ray missions (Table~\ref{tab:FoM}). In particular, also with their large FOVs, HUBS and Super-DIOS will be the most efficient instruments in detecting the extended hot CGM around local galaxies, although as small missions, their effective areas are not superb \citep{Cui19,Yamada18}. Angular resolution is not as important as in the study of point-like sources, because point source contributions (stellar sources, AGN) can always be well removed in spectral space, especially in narrow-band.


\subsection{Science with future X-ray missions}\label{subsec:NewScience}

With the significantly improved energy resolution in imaging spectroscopy mode, we expect to have some breakthrough on the study of the hot CGM with some future X-ray missions.

\begin{itemize}
\item \textbf{Nature of soft X-ray emission.} Diagnostic emission line features (e.g., the resonance, intercombination, and forbidden lines of the OVII~K$\alpha$ triplet) can only be resolved with high resolution X-ray spectra (e.g., with $E/\Delta E\gtrsim100$). These lines play a key role in determining the nature of the soft X-ray emission, i.e., if it is produced by thermal plasma or any non-thermal processes such as CX (e.g., \citealt{Zhang14}). The required resolution can be achieved with existing X-ray grating spectra (e.g., XMM-Newton/RGS) which has been less commonly adopted in the study of extended sources (the lines can be significantly broadened because of blending of emission from different parts of the extended source, e.g., Fig.~\ref{fig02}a). In the future, with more micro-calorimeter observations, we will be able to more accurately decompose the soft X-ray emission and determine its nature in more galaxies. 
\item \textbf{Metallicity of hot gas.} A reliable measurement of gas metallicity in X-ray in principle requires the separation of the lines and the continuum. In CCD spectra the lines and continuum are mixed together, so in spectral modeling the metallicity and gas density are degenerated and cannot be well constrained simultaneously. Most of the existing metallicity measurements are highly model dependent and uncertain, and usually only some abundance ratios are reliable (e.g., O/Fe; \citealt{Li15}). The condition can be significantly improved with micro-calorimeter.
\item \textbf{Dynamics of hot gas.} We need a velocity resolution of $\sim10^3\rm~km~s^{-1}$ (or $E/\Delta E\sim300$) to measure a line shift or broadening as small as $(200-300)\rm~km~s^{-1}$. This is the typical value of a moderate gas outflow (or global inflow) in a non-starburst galaxy, especially for the soft X-ray emitting gas which may be largely produced at the interface between the wind fluid and the entrained cool gas. We could thus be able to estimate the kinetic and turbulent energy contained in the hot CGM.
\item \textbf{Detection of extended CGM with narrow-band imaging.} In the study of hot CGM, we are mostly interested in a few diagnostic lines such the OVIII line at $\sim0.65\rm~keV$. In slightly redshifted galaxies, these lines can be separated from the same lines produced by the MW halo (the most important background component limiting the detection). Therefore, with high enough spectral resolution, we can conduct narrow-band imaging only including these lines, in order to achieve a much lower background and detection surface brightness. We did a simple simulation with the HUBS response files to show the basic concept of this method (Fig.~\ref{fig02}b,c). We also have a rough estimate of the requested energy resolution for a few local objects at different redshifts (Table~\ref{tab:NarrowBandImaging}). As the hot CGM is very faint and cannot be detected around too distant galaxies (e.g., with $d\gtrsim100\rm~Mpc$), an energy resolution of $(1-2)\rm~eV$ is required for the study of local galaxies at $d\lesssim20\rm~Mpc$ (e.g., the Virgo cluster galaxies). In this case, a large FOV (typically $\sim\rm degree$) is needed to cover a large fraction of the dark matter halo. This is why HUBS and Super-DIOS will be the best for narrow-band imaging observations of the hot CGM.
\end{itemize}

In addition to the significantly improved spectral resolution, the {\sl survey capability} of some missions (e.g., eROSITA) will also be very helpful in the study of hot CGM. This is not only for statistical analysis of large samples comparing the properties of different galaxy subsamples, but also help us to conduct stacking analysis probing the spatial distribution of faint extended hot CGM (e.g., \citealt{Anderson13,Li18}). Furthermore, we only discussed the X-ray emission line observations of the hot CGM, but the hot CGM can also be studied via {\sl absorption lines of background AGN}, or {\sl SZ effects}.

\begin{center}
\begin{table}[t]%
\centering
\caption{Parameters of Local Objects Which May Be Used for Narrow-band Imaging Observations.\label{tab:NarrowBandImaging}}%
\tabcolsep=0pt%
\begin{tabular*}{20pc}{@{\extracolsep\fill}lcccc@{\extracolsep\fill}}
\toprule
\textbf{Object} & \textbf{$d/\rm Mpc$} & \textbf{$v_\odot/\rm km~s^{-1}$} & $r_{\rm virial}/\rm deg$ & $\Delta E/\rm eV$ \\
\midrule
NGC891 & 10 & 530 & 1.5 & 1 \\
Virgo Cluster & 16 & 1280 & 4 & 2.5 \\
Fornax Cluster & 20 & 1340 & 1.9 & 2.7 \\
NGC5908 & 50 & 3300 & 0.5 & 6.5 \\
Perseus Cluster & 75 & 5400 & 1.3 & 10 \\
Coma Cluster & 100 & 6900 & 1.5 & 14 \\
\bottomrule
\end{tabular*}
\begin{tablenotes}
\item $v_\odot$ is the heliocentric receding velocity. $r_{\rm virial}$ is the apparent virial radius of the dark matter halo in degree. $\Delta E$ is the requested energy resolution to separate emission lines from the CGM of redshifted objects and the MW halo at $\sim0.6\rm~keV$.
\end{tablenotes}
\end{table}
\end{center}

\section*{Acknowledgments}

JTL acknowledge Dr. Shuinai Zhang to provide Fig.~\ref{fig02}a and the financial support from the \fundingAgency{National Aeronautics and Space Administration (NASA)} through the grants \fundingNumber{80NSSC19K0579}, \fundingNumber{80NSSC18K0536}, \fundingNumber{80NSSC19K1013}, and the \fundingAgency{Smithsonian Institution} through the grants \fundingNumber{AR9-20006X}.


\begin{thebibliography}{}

\bibitem[Anderson \& Bregman(2010)]{Anderson10} Anderson M. E., Bregman J. N., 2010, ApJ, 714, 320

\bibitem[Anderson \& Bregman(2011)]{Anderson11} Anderson M. E., Bregman J. N., 2011, ApJ, 737, 22

\bibitem[Anderson et al.(2013)]{Anderson13} Anderson M. E., Bregman J. N., Dai X., 2013, ApJ, 762, 106

\bibitem[Benson et al.(2000)]{Benson00} Benson A. J., Bower R. G., Frenk C. S., White S. D. M., 2000, MNRAS, 314, 557

\bibitem[Bogd\'{a}n et al.(2013)]{Bogdan13} Bogd\'{a}n A., Forman W. R., Vogelsberger M., et al., 2013, ApJ, 772, 97

\bibitem[Bogd\'{a}n et al.(2015)]{Bogdan15} Bogd\'{a}n A, Vogelsberger M., Kraft R. P., et al., 2015, ApJ, 804, 72

\bibitem[Bogd\'{a}n et al.(2017)]{Bogdan17} Bogd\'{a}n A, Bourdin H., Forman W. R., Kraft R. P., Vogelsberger M., Hernquist L., Springel V., 2017, ApJ, 850, 98

\bibitem[Bregman(2007)]{Bregman07} Bregman J. N., 2007, ARA\&A, 45, 221

\bibitem[Bregman et al.(2018)]{Bregman18} Bregman J. N., Anderson M. E., Miller M. J., et al., 2018, ApJ, 862, 3

\bibitem[Boroson et al.(2011)]{Boroson11} Boroson B., Kim D.-W., Fabbiano G., 2011, ApJ, 729, 12

\bibitem[Contursi et al.(2013)]{Contursi13} Contursi A., Poglitsch A., Graci\'{a} C. J., et al., 2013, A\&A, 549, 118

\bibitem[Crain et al.(2013)]{Crain13} Crain R. A., McCarthy I. G., Schaye J., Theuns T., Frenk C. S., 2013, MNRAS, 432, 3005

\bibitem[Cui et al.(2019)]{Cui19} Cui W., Chen L.-B., Gao B., et al., 2019, submitted to JLTP.

\bibitem[Dai et al.(2010)]{Dai10} Dai X., Bregman J. N., Kochanek C. S., Rasia E., 2010, ApJ, 719, 119

\bibitem[Dai et al.(2012)]{Dai12} Dai X., Anderson M., Bregman J., Miller J. M., 2012, ApJ, 755, 107

\bibitem[Faerman et al.(2017)]{Faerman17} Faerman Y., Sternberg A., McKee C. F., 2017, ApJ, 835, 52

\bibitem[Hitomi(2016)]{Hitomi16} Hitomi Collaboration, 2016, Nature, 535, 117

\bibitem[Hodges-Kluck et al.(2018)]{HodgesKluck18} Hodges-Kluck E. J., Bregman J. N., Li J.-t., 2018, ApJ, 866, 126

\bibitem[Kim \& Fabbiano(2013)]{Kim13} Kim D.-W., Fabbiano G., 2013, ApJ, 776, 116

\bibitem[Kim et al.(2019)]{Kim19} Kim D.-W., Anderson C., Burke D., et al., 2019, ApJS, 241, 36

\bibitem[Li \& Wang(2013a)]{Li13a} Li J.-T., Wang Q. D., 2013a, MNRAS, 428, 2085

\bibitem[Li \& Wang(2013b)]{Li13b} Li J.-T., Wang Q. D., 2013b, MNRAS, 435, 3071

\bibitem[Li et al.(2014)]{Li14} Li J.-T., Crain R. A., Wang Q. D., 2014, MNRAS, 440, 859

\bibitem[Li(2015)]{Li15} Li J.-T., 2015, MNRAS, 453, 1062

\bibitem[Li et al.(2016a)]{Li16a} Li J.-T., Beck R., Dettmar R.-J., et al., 2016a, MNRAS, 456, 1723

\bibitem[Li et al.(2016b)]{Li16b} Li J.-T., Bregman J. N., Wang Q. D., Crain R. A., Anderson M. E., 2016b, ApJ, 830, 134

\bibitem[Li et al.(2017)]{Li17} Li J.-T., Bregman J. N., Wang Q. D., Crain R. A., Anderson M. E., Zhang S., 2017, ApJS, 233, 20

\bibitem[Li et al.(2018)]{Li18} Li J.-T., Bregman J. N., Wang Q. D., Crain R. A., Anderson M. E., 2018, ApJL, 855, 24

\bibitem[Liu et al.(2011)]{Liu11} Liu J., Mao S., Wang Q. D., 2011, MNRASL, 415, 64

\bibitem[Liu et al.(2012)]{Liu12} Liu J., Wang Q. D., Mao S., 2012, MNRAS, 420, 3389

\bibitem[McCammon(2005)]{McCammon05} McCammon D., Cryogenic Particle Detection, Topics in Applied Physics, v99. Springer-Verlag Berlin/Heidelberg, 2005, p1 

\bibitem[Miller \& Bregman(2015)]{Miller15} Miller M. J., Bregman J. N., 2015, ApJ, 800, 14

\bibitem[Mineo et al.(2014)]{Mineo14} Mineo S., Gilfanov M., Lehmer B. D., Morrison G. E., Sunyaev R., 2014, MNRAS, 437, 1698

\bibitem[Mora-Partiarroyo et al.(2019)]{MoraPartiarroyo19} Mora-Partiarroyo S. C., Krause M., Basu A., et al., 2019, submitted to A\&A

\bibitem[O'Sullivan et al.(2003)]{OSullivan03} O'Sullivan E., Ponman T. J., Collins R. S., 2003, MNRAS, 340, 1375

\bibitem[Otte et al.(2003)]{Otte03} Otte B., Murphy E. M., Howk J. C., Wang Q. D., Oegerle W. R., Sembach K. R., 2003, ApJ, 591, 821

\bibitem[Ranalli et al.(2008)]{Ranalli08} Ranalli P., Comastri A., Origlia L., Maiolino R., 2008, MNRAS, 386, 1464

\bibitem[Rasmussen et al.(2009)]{Rasmussen09} Rasmussen J., Sommer-Larsen J., Pedersen K., Toft S., Benson A., Bower R. G., Grove L. F., 2009, ApJ, 697, 79 

\bibitem[Revnivtsev et al.(2009)]{Revnivtsev09} Revnivtsev M., Sazonov S., Churazov E., Forman W., Vikhlinin A., Sunyaev R., 2009, Nature, 458, 1142

\bibitem[Smith et al.(2012)]{Smith12} Smith R. K., Foster A. R., Brickhouse N. S., 2012, AN, 333, 301

\bibitem[Strickland et al.(2002)]{Strickland02} Strickland D. K., Heckman T. M., Weaver K. A., Hoopes C. G., Dahlem M., 2002, ApJ, 568, 689

\bibitem[Strickland et al.(2004a)]{Strickland04a} Strickland D. K., Heckman T. M., Colbert E. J. M., Hoopes C. G., Weaver K. A., 2004a, ApJS, 151, 193

\bibitem[Strickland et al.(2004b)]{Strickland04b} Strickland D. K., Heckman T. M., Colbert E. J. M., Hoopes C. G., Weaver K. A., 2004b, ApJ, 606, 829

\bibitem[Strickland \& Heckman(2007)]{Strickland07} Strickland D. K., Heckman T. M., 2007, ApJ, 658, 258

\bibitem[Strickland \& Heckman(2009)]{Strickland09} Strickland D. K., Heckman T. M., 2009, ApJ, 697, 2030

\bibitem[Strong et al.(2010)]{Strong10} Strong A. W., Porter T. A., Digel S. W., et al., 2010, ApJL, 722, 58

\bibitem[Sun et al.(2009)]{Sun09} Sun M., Voit G. M., Donahue M., et al., 2009, ApJ, 693, 1142

\bibitem[Tang et al.(2009)]{Tang09} Tang S. K., Wang Q. D., Lu Y., Mo H. J., 2009, MNRAS, 392, 77

\bibitem[Toft et al.(2002)]{Toft02} Toft S., Rasmussen J., Sommer-Larsen J., Pedersen K., 2002, MNRAS, 335, 799

\bibitem[T\"{u}llmann et al.(2006a)]{Tullmann06a} T\"{u}llmann R., Pietsch W., Rossa J., Breitschwerdt D., Dettmar R.-J., 2006a, A\&A, 448, 43

\bibitem[T\"{u}llmann et al.(2006b)]{Tullmann06b} T\"{u}llmann R., Breitschwerdt D., Rossa J., Pietsch W., Dettmar R.-J., 2006b, A\&A, 457, 779

\bibitem[Vikhlinin et al.(2006)]{Vikhlinin06} Vikhlinin A., Kravtsov A., Forman W., et al., 2006, ApJ, 640, 691

\bibitem[Wang(2010)]{Wang10} Wang Q. D., 2010, PNAS, 107, 7168

\bibitem[Wang et al.(2016)]{Wang16} Wang Q. D., Li J., Jiang X., Fang T., 2016, MNRAS, 457, 1385

\bibitem[Yamada et al.(2018)]{Yamada18} Yamada S., Ohashi T., Ishisaki Y., et al., 2018, JLTP, 193, 1016 

\bibitem[Zhang et al.(2014)]{Zhang14} Zhang S., Wang Q. D., Ji L., Smith R. K., Foster A. R., Zhou X., 2014, ApJ, 794, 61

\end{thebibliography}
\end{document}